\documentclass[]{IEEEtran}

\input epsf
\usepackage{graphicx}
\usepackage{epsfig}
\usepackage[tight,hang]{subfigure}

\usepackage{units}
\usepackage[utf8]{inputenc}	
\usepackage{amsmath}
\usepackage{amssymb}
\usepackage{upgreek}

\begin{document}

\title{\LARGE Thin-Film InGaAs Metamorphic Buffer for telecom C-band InAs Quantum Dots and Optical Resonators on GaAs Platform
}
\author{R. Sittig*, C. Nawrath, S. Kolatschek, S. Bauer, R. Schaber,\\ 
J. Huang, P. Vijayan, S. L. Portalupi, M. Jetter and P. Michler\\
Institut f\"ur Halbleiteroptik und Funktionelle Grenzfl\"achen,\\ Center for Integrated Quantum Science and Technology (IQ$^{ST}$)\\ and SCoPE,\\ University of Stuttgart,\\ Allmandring 3, 70569 Stuttgart, Germany\\E-Mail: r.sittig@ihfg.uni-stuttgart.de}

\maketitle

\begin{abstract}
The GaAs-based material system is well-known for the implementation of InAs quantum dots (QDs) with outstanding optical properties  \cite{michler2017quantum}.
However, these dots typically emit at a wavelength of around \unit[900]{nm}.  
The insertion of a metamorphic buffer (MMB)
can shift the emission to the technologically attractive telecom C-band range centered at \unit[1550]{nm}.
However, the thickness of common MMB designs limits their compatibility with most photonic resonator types \cite{senellart2017high}.
Here we report on the MOVPE growth of a novel InGaAs MMB with a non-linear indium content grading profile designed to maximize plastic relaxation within minimal layer thickness. 
Single-photon emission at \unit[1550]{nm} from InAs QDs deposited on top of this thin-film MMB is demonstrated.  
The strength of the new design is proven by integrating it into a bullseye cavity via nano-structuring techniques.
The presented advances in the epitaxial growth of QD/MMB structures 
form the basis for the fabrication of high-quality telecom non-classical light sources as a key component of photonic quantum technologies.
\end{abstract}
\section*{Motivation} 
There are two main approaches to reduce the blue-shifting strain on InAs QDs and obtain emission at \unit[1550]{nm}.
The first is a change of the substrate from GaAs to InP \cite{takemoto2004observation,skiba2017universal}, 
and the second, the growth of a MMB below the QDs \cite{ledentsov2007mbe,semenova2008metamorphic}.
Single-photon emission from InAs QDs grown on InP has been studied extensively for decades 
and significant advances have been reported in recent years \cite{muller2018quantum,anderson2020gigahertz}.
Nevertheless, from a growth perspective, the MMB on GaAs approach offers distinct benefits: it avoids the introduction of phosphorous compounds into the structure and allows to fine-tune the strain via the lattice constant of the matrix, providing more freedom for the QD growth parameters. Additionally, it unlocks access to efficient binary AlAs/GaAs distributed Bragg reflectors (DBRs) as well as AlGaAs-based  etch-stop and sacrificial layers for various processing techniques.\\
MMBs consisting of III-V materials are well-established in a wide range of semiconductor devices, 
such as high electron mobility transistors \cite{ajayan2019gaas} and multi-junction solar cells \cite{philipps2012present}. The functionality of these devices is generally independent of the MMB thickness, provided that a high crystalline quality is ensured.
Therefore, MMBs are typically at least \unit[1]{$\upmu$m} thick and their material composition is linearly or step-graded, because this approach facilitates control over properties like surface roughness and defect density  \cite{sorokin2016peculiarities}.\\
Likewise, a linear InGaAs MMB with a thickness of \unit[1080]{nm} enabled the first demonstration of single-photon emission in the telecom C-band from InAs QDs grown on the GaAs material platform \cite{paul2017single}.
Furthermore, polarization-entanglement \cite{olbrich2017polarization}, on–demand generation of entangled, single photon pairs \cite{zeuner2019demand} and indistinguishability under cw two-photon-resonant \cite{nawrath2019coherence} and pulsed resonant \cite{nawrath2021resonance} excitation was shown by utilizing the same design. 
However, this sample structure features only a nominal $3\uplambda$-cavity between bottom DBR and semiconductor/air interface.
Employing an advanced photonic structure, e.g. a $\uplambda$-cavity micro-pillar, 
would substantially improve the emission properties \cite{senellart2017high}.
Therefore, enabling device fabrication is a crucial next step.\\ 
A suitable MMB design must fulfil three mandatory requirements.
First, provide sufficient strain reduction to shift the QD emission to \unit[1550]{nm}.
Second, a smooth and homogeneous surface is a necessary prerequisite for all processing techniques.
Third, in order to retain its compatibility with the AlAs/GaAs material system,
the MMB must be placed inside the $\uplambda$-cavity for most photonic structures. This puts a strict upper limit on its thickness. Consequently, the previous linear design is unsuitable for this purpose and a thin-film replacement has to promote an extremely efficient transition of the lattice constant, 
while maintaining a high crystalline quality.
 \section*{Relaxation-optimized Buffer Design}
\begin{figure}
	\centering
	\includegraphics[width=\columnwidth]{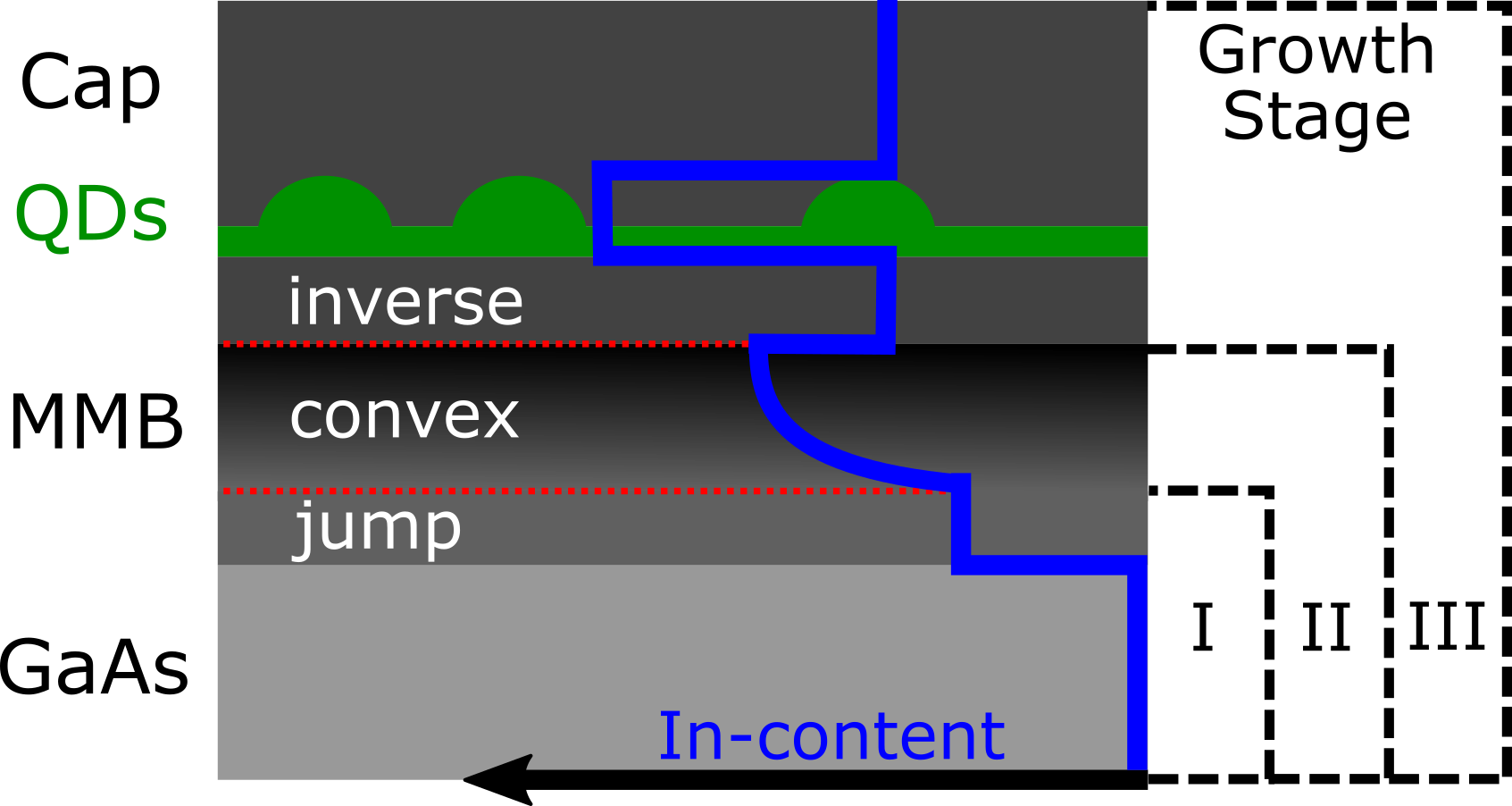}
	\caption{Schematic structure of InAs QDs 
		inside an InGaAs-matrix lattice matched to a
		MMB grown on a GaAs substrate. 
		The varying indium content throughout the structure is illustrated in blue.
		Growth optimization is performed in three consecutive growth stages I, II and III. 		
	}
	\label{SchematicJCI}
\end{figure}

The intentional alteration of the lattice constant during metamorphic heteroepitaxy is achieved by inducing the formation of misfit dislocation segments into the previously pseudomorphic layer. Thus, the strained layer relaxes and adopts an in-plane lattice constant closer to its inherent value.  
Although the exact mechanisms responsible for the formation of these segments are only partly understood, this process is clearly driven by strain energy  
\cite{matthews1970accommodation,kujofsa2012plastic}. 
Therefore, a large lattice mismatch with the substrate 
is desirable for the growth of a thin MMB.
However, there is an upper limit for the possible mismatch, which is given by the onset of 3D-growth as a competing mechanism of strain energy reduction.
Our proposed content grading profile
to provide the maximum possible strain energy at any point, 
while staying within the limits of 2D-growth, 
is depicted by the InGaAs MMB in Fig.\,\ref{SchematicJCI}.  
The grading begins with an abrupt change (jump) of the indium content, followed by a convex-up grading, and is completed by an inverse step.  
This jump-convex-inverse design is optimized in three successive growth stages named I, II and III. \\
In stage I, which corresponds to only the jump-layer, the aim is to induce a quick start of the plastic relaxation. 
Thus, we need to find the maximal permissible In content for this jump and the minimal thickness for the onset of relaxation. \\
Once the lattice constant begins to increase, the indium content
can be increased further without generating 3D-structures.
Following the same rationale, 
a convex-up function, in analogy to the metamorphic relaxation curve \cite{andrews2002modeling}, is deduced as the steepest possible grading profile.
This jump into convex-up design is compared with several alternatives in growth stage II.
Additionally, the minimum thickness (i.e. maximum grading) that still maintains low surface roughness has to be determined.
Furthermore, in order to reach the desired lattice constant, but avoid the stagnant saturation regime \cite{rodriguez2004role}, an overshoot of In content is employed in the convex region before a decrease in the inverse region (see the dip at interface convex/inverse).
Notably, this overshoot has to be adjusted retroactively to enable a lattice-matched deposition of the fully-relaxed inverse layer.\\
The main role of the inverse layer is to serve as a substrate for the QD deposition and (together with the capping layer) to provide the correct amount of strain release
to allow for the formation of InAs QDs with emission within the telecom C-band. 
This will therefore be the pivotal criterion for the optimization in stage III.\\
The key steps of the MMB growth calibration and optimization are presented in the following.
\section*{Optimization Procedure}
 \subsection*{Jump layer optimization}
 \begin{figure}
 	\centering
 	\includegraphics[width=\columnwidth]{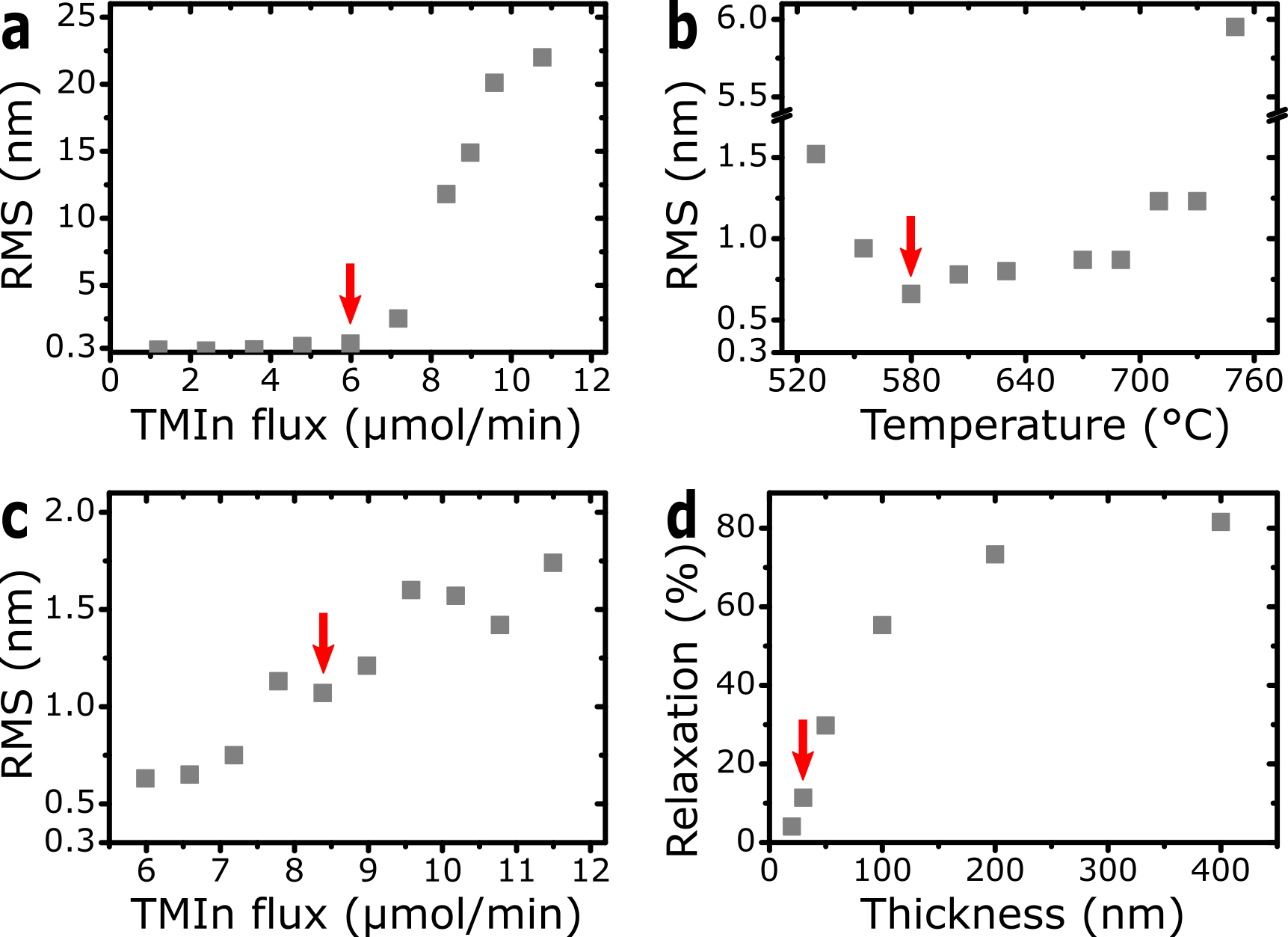}
 	\caption{Successive optimization steps of growth stage I.
 			The red arrows mark the obtained parameter values.
 		a-c: RMS surface roughness of InGaAs layers with varying growth parameters. 				
 		a: Variation of TMIn flux for the growth of \unit[50]{nm} InGaAs with \unit[20.8]{$\upmu$mol/min} TMGa flux and \unit[2973]{$\upmu$mol/min} AsH$_3$ flux at \unit[710]{$^{\circ}$C}.
 		b: Temperature dependence of InGaAs with a TMIn flux of \unit[6.0]{$\upmu$mol/min}. 
 		c: Variation of TMIn flux at \unit[580]{$^{\circ}$C}. 
 		d: Strain relaxation of In$_{27.4}$GaAs depending on layer thickness.
	}
 	\label{Optimization_j}
 \end{figure} 
We adopted the starting parameters for the optimization of the jump-layer (stage I)
from our default metal-organic vapor-phase epitaxy (MOVPE) growth of high quality GaAs, namely a temperature of \unit[710]{$^\circ$C}, a TMGa flux of \unit[20.8]{$\upmu$mol/min} and an AsH$_3$ flux of \unit[2973]{$\upmu$mol/min}.
We then grew samples with \unit[50]{nm} thick layers of InGaAs by adding varying amounts of TMIn into the mix.
As shown in Fig.\,\ref{Optimization_j}a, the RMS surface roughness of these layers exhibits a sharp rise for higher TMIn fluxes, indicating a transition towards 3D growth. We therefore selected to proceed with a value of \unit[6]{$\upmu$mol/min} for further investigation.\\
The next step was to examine the influence of the growth temperature, as this parameter is crucial for controlling the diffusion of atoms on the surface and the mobility of dislocations inside the layer \cite{sasaki2011growth}.
A corresponding comparison of surface roughness is displayed in Fig.\,\ref{Optimization_j}b and
establishes a temperature of \unit[580]{$^\circ$C} as advantageous.\\
Next, we reiterated the determination of the maximum possible indium content, because a lower temperature is expected to suppress 3D-growth  \cite{ceschin1991strain}.
In contrast to the results at \unit[710]{$^\circ$C}, no clear transition can be identified, instead we observe an approximately linear relation between RMS and TMIn flux as shown in Fig.\,\ref{Optimization_j}c.
However, the  surface topography for higher indium contents, reveals ordering along the diagonal $[001]$ and $[010]$ directions (see Supplementary).
In contrast, metamorphic layers typically exhibit a $[011]$/$[0\bar{1}1]$ cross-hatch pattern \cite{andrews2002modeling}. 
Therefore, we decided to avoid this regime and adopted \unit[8.4]{$\upmu$mol/min} 
as the maximum applicable TMIn flux for the jump layer. \\
With the material composition and the temperature defined, the next step was to find the minimum thickness at which the layer starts to relax.
For this purpose, we grew samples with InGaAs layer thickness ranging from \unit[20]{nm} to \unit[400]{nm} and determined their respective relaxation via X-ray diffraction.
 The results are displayed in Fig.\,\ref{Optimization_j}d.
The relaxation curve exhibits the typical behavior of metamorphic growth, namely a steep increase after a certain critical thickness, followed by saturation for thicker layers.
 \unit[30]{nm} is determined as the minimal thickness that exhibits a clear onset of relaxation (\unit[11.4]{\%})
and is therefore used as parameter for the jump-step layer.
This completes the optimization of growth stage I.
 \subsection*{Convex grading layer optimization}
 \begin{figure}
 	\centering
 	\includegraphics[width=\columnwidth]{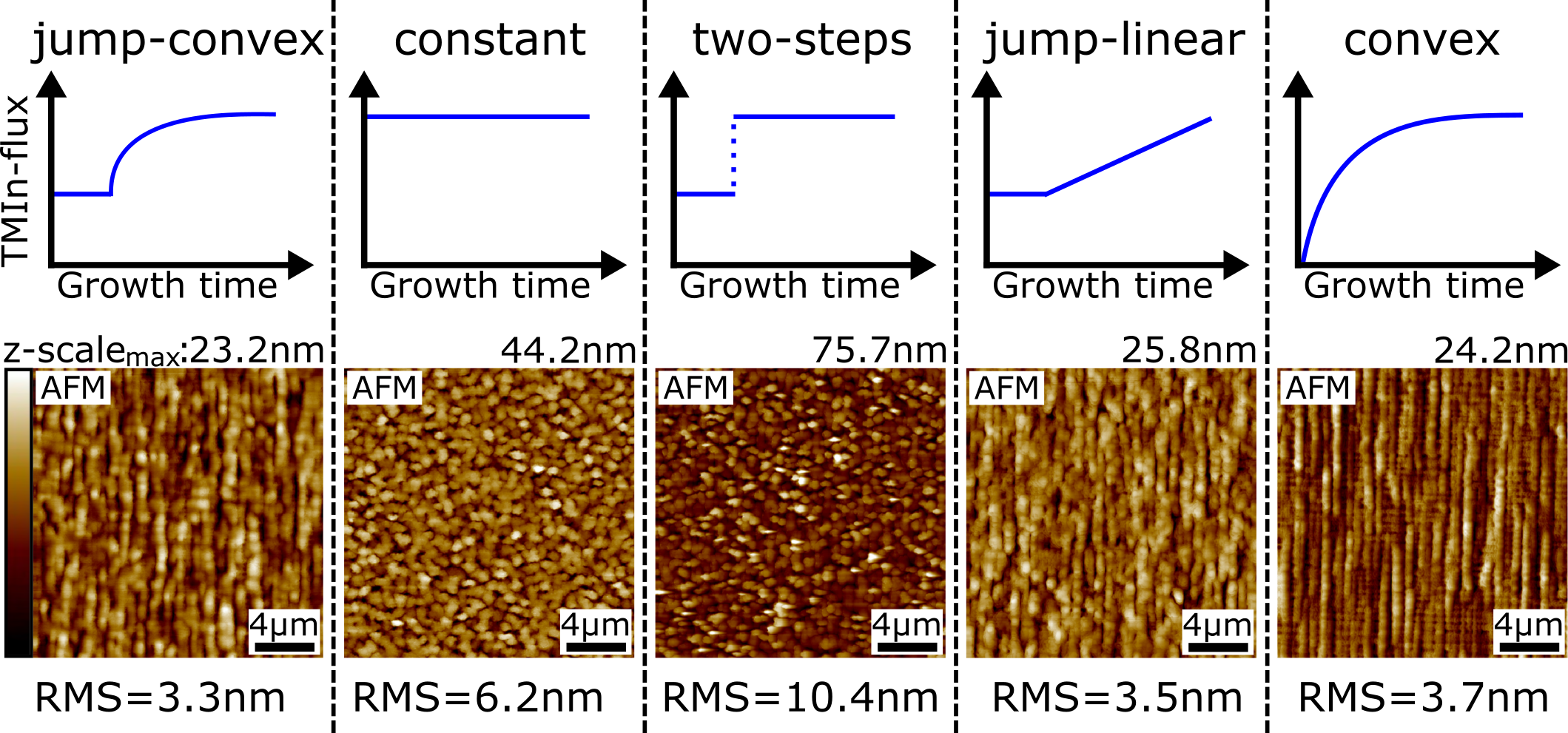}
 	\caption{Optimization of growth stage II:		 
 		Comparison between AFM-scans of various In$_x$Ga$_{1-x}$As grading profiles to reach \unit[13.2]{$\upmu$mol/min} TMIn flux.		
 		 The constant section used in profiles 1, 3 and 4 has a thickness of \unit[30]{nm} and the total thickness of each structure is \unit[200]{nm}.}	
 	\label{Optimization_c}
 \end{figure} 
Fig.\,\ref{Optimization_c} shows a comparison of our proposed jump-convex content-grading profile with various alternatives, each reaching a maximum TMIn flux of \unit[13.2]{$\upmu$mol/min} within \unit[200]{nm}. 
Using a constant or two-steps profile is clearly inferior due to their higher surface roughness.
Furthermore, employing a jump-linear or convex profile results in a similar RMS, but provides less average strain for the relaxation process.   
Therefore, we proceeded with finding a suitable thickness for the grading layer.
Notably, a thinner convex segment
steepens the necessary grading not only directly, but also indirectly by decreasing the final relaxation,
which entails a larger overshoot to reach the desired effective lattice constant.
\unit[130]{nm} was determined as suitably thin, but still allowed adjusting the final indium content without 
inducing 3D-growth. This is necessary for the next task 
 of simultaneously fine-tuning the composition of convex and inverse/capping layer to obtain QD emission at \unit[1550]{nm} in growth stage III.  
 \subsection*{Inverse layer optimization} 
\begin{figure}
	\centering
	\includegraphics[width=0.95\columnwidth]{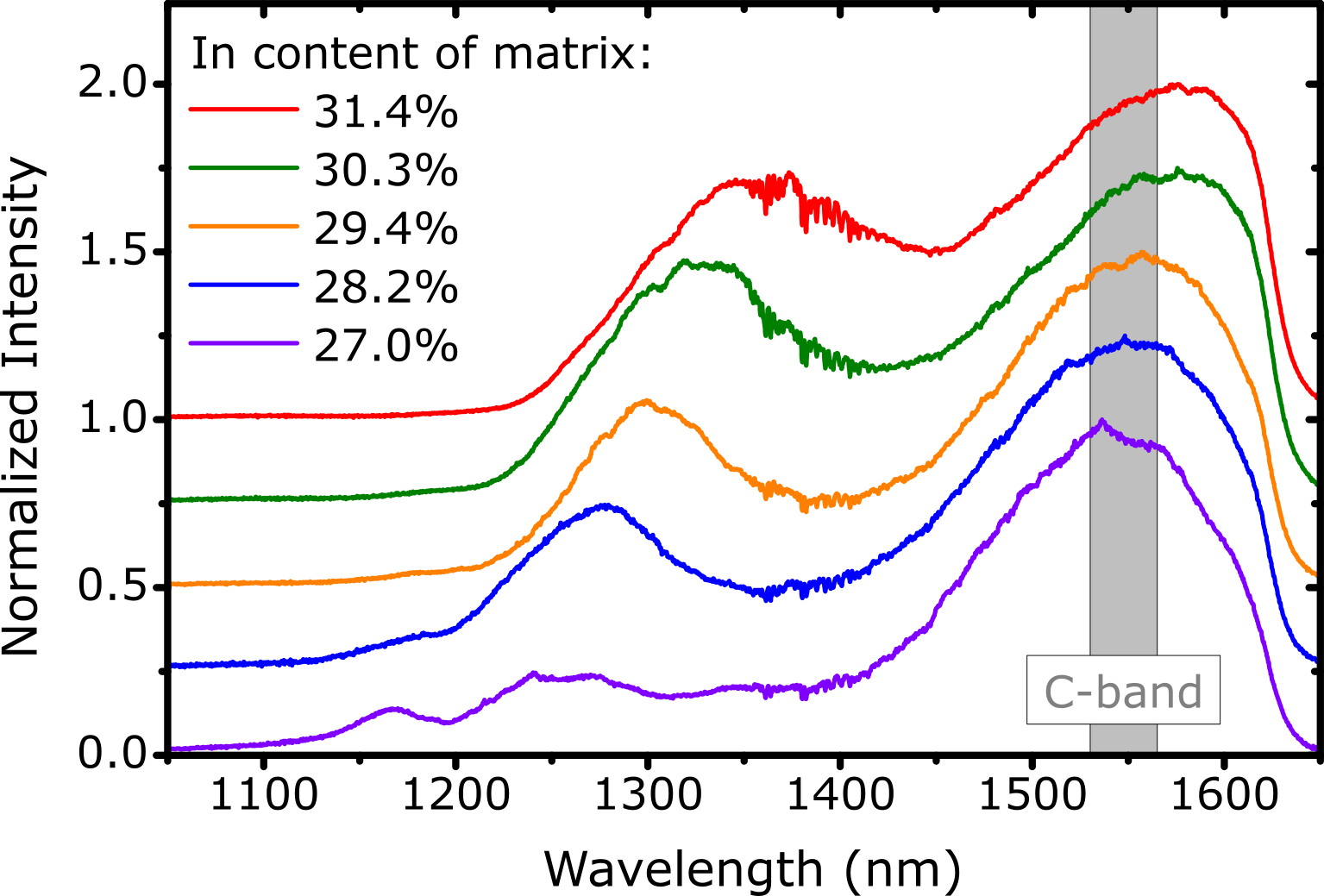}
	\caption{Optimization of growth stage III: Photoluminescence spectra of samples with varying indium content in the In$_{x}$Ga$_{1-x}$As matrix around the QDs. The curves are offset vertically for clarity.}
	\label{Optimization_i}
\end{figure}

For this purpose, we fabricated five capped samples with the indium content inside the matrix around the QDs ranging between \unit[27.0]{\%} and \unit[31.4]{\%} (\unit[35.9]{\%} to \unit[40.6]{\%} at maximum of convex layer).
The thickness of the inverse/capping layer was set to \unit[60]{nm}/\unit[220]{nm}.
This sets the QDs in the center of a structure with a total thickness of \unit[440]{nm},
which corresponds to the approximate geometrical length of a $\uplambda$-cavity consisting of 
the deposited InGaAs \cite{goldberg1999handbook}.
A comparison between the photoluminescence spectra of the five samples is shown in Fig.\,\ref{Optimization_i}.
Each spectrum is composed of two peaks. The peak around \unit[1550]{nm} can be identified as QD emission and the one at shorter wavelengths is assumed to
stem from the wetting layer or the InGaAs matrix.
As expected, the higher the In content of the matrix, the stronger the red-shift of the emission.
Microphotoluminescence ($\upmu$-PL) measurements (not shown) revealed that the In$_{29.4}$GaAs configuration produces the highest percentage of 
 single QDs emitting in the \unit[1545-1555]{nm} range. 
Hence, we selected this indium content as the most suitable, which finalized the optimization of the jump-convex-inverse design. 
\section*{Photonic Structures}
 \begin{figure*}
	\centering
	\includegraphics[width=\textwidth]{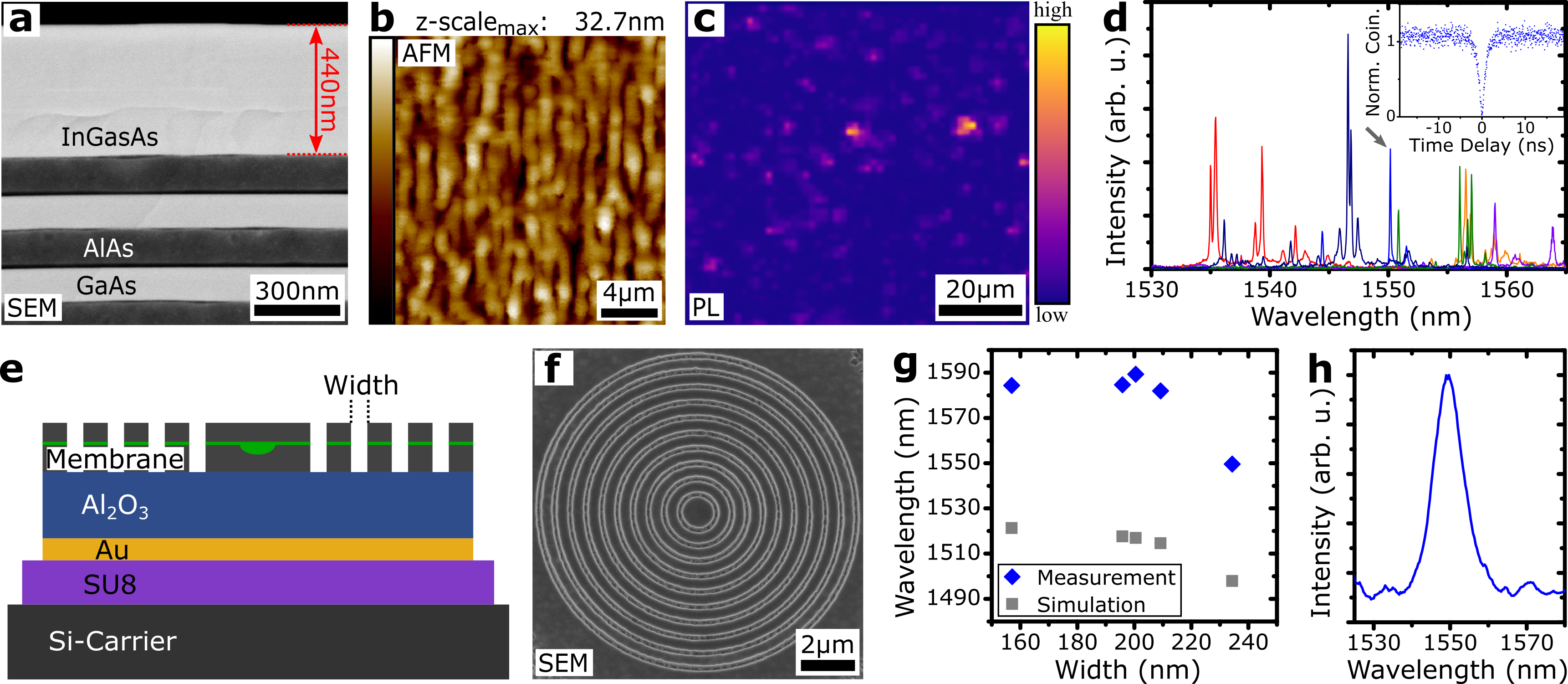}
	\caption{a-d: Structural and optical investigation of a QD sample featuring an In$_{29.4}$GaAs-matrix and an AlAs/GaAs DBR. 
		a: SEM side view of the two upper DBR-pairs, the jump-convex-inverse MMB and the capped QDs.	 
		b: AFM scan showing the surface topography. The extracted RMS is \unit[4.5]{nm}. 	
		c: $\upmu$-PL map.  The	color scale represents the emission intensity inside the C-band.
		An area density for the QDs of 5 x \unit[10$^6$]{cm$^{-2}$} was extracted. 
		d: $\upmu$-PL spectra of six exemplary QDs emitting inside the telecom C-band. Inset: Autocorrelation measurement of the blue transition line (marked with a grey arrow) under continuous wave above band-gap excitation.	
		e-h: Bullseye structure for emission inside the telecom C-band.
		e: Schematic of the sample after the flip-chip process. An Al$_2$O$_3$ spacer layer between the membrane and the backside gold mirror is used to enable constructive interference between the upwards and the backreflected emission.
		f: SEM top view of an exemplary processed circular Bragg grating cavity.
		g: Simulated and measured resonance wavelength of the fundamental cavity mode as a function of the trench width. 
		h: $\upmu$-PL spectrum of the cavity peak at \unit[1550]{nm} under high power above band-gap pumping.
	}
	\label{DBR_Bullseye}
\end{figure*}
In order to illustrate the potential and quality of the novel MMB we fabricated two exemplary structures.\\
The first one was grown on top of 23 AlAs/GaAs DBR  pairs for increased light extraction.
The SEM scan in Fig.\,\ref{DBR_Bullseye}a displays a side view of the two upper DBR pairs, the MMB and the capped QDs.
The InGaAs structure is \unit[440]{nm} thick in total, as discussed above. 
Notably, growth on AlAs instead of GaAs had no adverse influence on the quality of the MMB.
The AFM scan in Fig.\,\ref{DBR_Bullseye}b displays the surface topography of the sample.
It is dominated by a cross-hatch pattern, which is typical for MMB structures \cite{andrews2002modeling}, resulting in an RMS of \unit[4.5]{nm}.
This topography remains homogeneous over the whole sample.
Fig.\,\ref{DBR_Bullseye}c shows a  $\upmu$-PL map of the sample in which the color scale depicts the emission intensity inside the telecom C-band. 
An analysis of distinct emitters allows to extract an area density for the QDs of $\sim$5 x \unit[10$^6$]{cm$^{-2}$},
which signifies excellent conditions for the excitation of isolated emitters, 
as required in quantum optical experiments.
A selection of exemplary $\upmu$-PL spectra acquired under non-resonant excitation is  shown  in Fig.\,\ref{DBR_Bullseye}d.     
The displayed emission consists of multiple sharp transition peaks,
which can be found over the full range of the telecom C-band and are characteristic for single QDs.
The non-classical nature of this emission
is confirmed by the anti-bunching dip in the second-order auto-correlation function displayed in inset Fig.\,\ref{DBR_Bullseye}d.   
At zero time delay a raw value of $g^{(2)}(0)=0.026(1)$   
is found.\\
The second structure is a circular Bragg grating (Bullseye) cavity, that was realized by combining the thin-film MMB with an AlGaAs sacrificial layer. This enables the use of a flip-chip process similar to Ref.~\cite{liu2019solid,wang2019demand} resulting in a layer structure as sketched in Fig.\,\ref{DBR_Bullseye}e. 
Reducing the inverse/capping layer thickness to \unit[20]{nm}/\unit[180]{nm}
produces a \unit[360]{nm} thick membrane, that allows for sufficient confinement in the  growth direction via total internal reflection.  
The SEM image in Fig.\,\ref{DBR_Bullseye}f shows a processed Bullseye structure after electron-beam lithography and chemical dry etching.
$\upmu$-PL measurements were performed to investigate the cavity mode,
 ensuring its uniform feeding by high power above band-gap pumping.
The spectral position of the fundamental mode
for different trench widths is plotted in Fig.\,\ref{DBR_Bullseye}g.
For smaller values, a saturation behaviour of the resonance wavelength is observed, while for increasing trench widths the cavity mode starts to blue-shift. 
This trend is verified by finite-difference time-domain  
simulations. However, a small offset between simulations and measurements 
is observed. This can possibly be explained by the simplified and averaged refractive index used in the simulations. In contrast, the real refractive index is highly dependent on the material composition and temperature. Furthermore, a small variation of the membrane or Al$_2$O$_3$ layer thickness can yield an additional small difference. 
An exemplary spectrum for a trench width of \unit[240]{nm} with a cavity mode around \unit[1550]{nm} is displayed in Fig.\,\ref{DBR_Bullseye}h, thus demonstrating the feasibility of fabricating Bullseye cavities in the telecom C-band based on the jump-convex-inverse MMB design.

\section*{Conclusion and Outlook} 
We realized the MOVPE growth of
a thin-film ($\leq$\unit[220]{nm}) InGaAs MMB on a GaAs substrate, by optimizing a non-linear grading profile designed for efficient plastic relaxation.
Furthermore, we demonstrated that the strain on InAs QDs placed inside a respective InGaAs matrix can be adjusted to obtain telecom C-band photons.
The MMB/QD structure was subsequently grown directly on a GaAs/AlAs DBR. 
This sample exhibits high surface quality and allowed us to demonstrate single-photon emission.
Finally, the structure was integrated into a Bullseye cavity to prove its compatibility with highly appealing photonic resonator designs. 
This progress in MMB growth unlocks numerous approaches for the fabrication of 
high quality single-photon sources at \unit[1550]{nm}.

\section*{Acknowledgements}
 We gratefully acknowledge the funding from the German Federal Ministry of Education and Research (BMBF) via
 the project Q.Link.X (No. 16KIS0862). We further acknowledge the European Union’s  Horizon 2020 research and innovation program under Grant Agreement No. 899814 (Qurope). Partial funds of this work were provided by Projects No. EMPIR 17FUN06 SIQUST and No. 20FUN05 SEQUME. These projects have received funding from the EMPIR programme co-financed by the Participating States and from the European Union’s Horizon 2020 research and innovation program.

\section*{Methods} 
The structures were grown in a commercial AIX-200 horizontal metal-organic vapor-phase epitaxy (MOVPE) system at a pressure of 100 mbar using the standard
precursors trimethylgallium (TMGa), trimethylindium (TMIn), trimethylaluminum (TMAl) and arsine (AsH$_3$).
We used exactly oriented (100) GaAs substrates and deposited a GaAs buffer to provide a high surface quality for the subsequent layers.\\
The compositional change within the In$_x$Ga$_{1-x}$As MMB in this work was realized by fixing the TMGa flux and adjusting the TMIn flux.
This leads to a minor distortion of the grading profiles, because the deposition speed is slightly increased at higher indium contents (approx. \unit[17]{\%} increase between In$_{27.5}$GaAs and In$_{38}$GaAs).
However, this should have a negligible impact on the conclusions made for the comparison between different profiles.
QDs were grown by depositing InAs for \unit[4]{s} with a TMIn-flux of \unit[9.6]{$\upmu$mol/min} at \unit[595]{$^\circ$C} independent of the specific structure or sample.\\ 
Atomic force microscopy (AFM) was used to investigate
the surface topology and to determine its roughness given by the root mean square (RMS). 
This was our main criterion for the crystal quality of the MMB layers, because surface roughness prominently influences QD growth, excitation/extraction efficiency and the compatibility with nano-structuring techniques.
All scans are depicted with the horizontal axis being aligned to the intersection of the surface with the Ga/In-facets ($[011]/[0\bar{1}\bar{1}]$) and the vertical axis being aligned to the respective As-facets ($[0\bar{1}1]$/$[01\bar{1}]$).\\
We recorded reciprocal space maps (RSM) in a high-resolution X-ray diffractometer (XRD) 
to determine the composition and relaxation inside MMB layers with a method similar to Ref.~\cite{chauveau2003indium}.\\
Deposition rates were calibrated by measuring layer thicknesses  
with a combination of scanning electron microscopy (SEM) and X-ray reflectivity (XRR).\\
In order to investigate the QD emission characteristics, photoluminescence (PL) or micro-photoluminescence  ($\upmu$-PL) measurements were performed 
at \unit[4]{K}, depending if an ensemble of QDs or a single QD is addressed, respectively. 
The dots were optically excited by a continuous wave (cw) laser above the bandgap of the surrounding matrix material. The QD emission is collected by the same objective 
used to focus the excitation laser and guided into a spectrometer. To measure the second-order auto-correlaction function $g^{(2)}(\tau)$, a single QDs transition is filtered with a spectral width of $\approx$\unit[5]{GHz} and coupled to a fiber-based Hanbury-Brown and Twiss type setup consisting of a beam splitter and two superconducting nanowire single-photon detectors. The full detection system including the subsequent time tagging electronics exhibit a temporal resolution of $\approx$\unit[45]{ps}.

\end{document}